\documentclass[preprint,showpacs,aps]{revtex4}

\usepackage{amsmath}
\usepackage{dcolumn}
\usepackage{verbatim}
\usepackage{mathrsfs}
\usepackage{graphicx}

\begin{document}

\title{Full major-shell calculation for states that were degenerate in a 
single-$j$-shell calculation}

\author{A. Escuderos$^1$, S.J.Q.~Robinson$^2$, and L.~Zamick$^1$}

\affiliation{$^1$Department of Physics and Astronomy, Rutgers University,
Piscataway, New Jersey 08854}
\affiliation{$^2$Geology and Physics Department, University of Southern 
Indiana, Evansville, Indiana 47712}

\date{\today}

\begin{abstract} 
A full $fp$ calculation is performed for states which were degenerate in a 
single-$j$-shell calculation in which isospin-zero two-body matrix elements 
were set to zero energy. Most of the splitting in a complete shell calculation 
(but not all) comes from the $T=0$ part of the interaction.
\end{abstract}

\pacs{}

\maketitle

In a previous work~\cite{ebzr05}, we 
explained why certain states were degenerate in the 
single $j$ shell for an interaction in which the isospin $T=0$ two-body matrix 
elements were set to zero. The degeneracy splitting was also obtained by 
reintroducing the full interaction. In this work, we shall calculate the energy
splittings in a full $fp$-shell calculation.

In our single-$j$-shell calculation with $j=f_{7/2}$, we took the two-body 
matrix elements $E(J)=\langle (f_{7/2}^2)^J |V| (f_{7/2}^2)^J \rangle$ from 
experiment, i.e., from the spectrum of $^{42}$Sc. The even-$J$ states of the 
$f_{7/2}^2$ configuration have isospin $T=1$, while the odd-$J$ states have 
isospin $T=0$. The values of $E(J)$ in MeV (with the $J=0$ state taken to be at
zero energy) are:
\begin{center}
\begin{tabular*}{.5\textwidth}{@{\extracolsep{\fill}}rrcrr}
\multicolumn{2}{c}{$T=1$} & & \multicolumn{2}{c}{$T=0$} \\ 
\cline{1-2} \cline{4-5}
$J=0$ & 0.0000 & & $J=1$ & 0.6111 \\
$J=2$ & 1.5863 & & $J=3$ & 1.4904 \\
$J=4$ & 2.8153 & & $J=5$ & 1.5101 \\
$J=6$ & 3.2420 & & $J=7$ & 0.6163 \\
\end{tabular*}
\end{center}

The interaction $T0E(J)$ is obtained by setting the second column to zero. 
There will actually be no difference if we set the $T=0$ matrix elements to a 
constant, as long as we are considering splittings of states of the same 
isospin, which indeed we are.

In the single-$j$-shell case, the degeneracies fall into two classes. In the 
first case, we have degeneracies of states in $^{43}$Sc ($^{43}$Ti) and 
$^{44}$Ti. For these, we have an explanation---a partial dynamical 
symmetry~\cite{rz0163,rz0164}. We found that the degeneracies for $T0E(J)$ 
occurred for states with angular momenta that could not exist for systems of 
identical particles, i.e., in $^{43}$Ca and $^{44}$Ca, respectively. In the 
second case, we have all other degeneracies that are listed in 
Table~\ref{tab:full}, selected states in $^{45}$Ti, $^{46}$V, and 
$^{47}$V~\cite{ebzr05}. For these, we could not find any symmetries related to 
the degeneracies and concluded that we indeed had degeneracy without symmetry.

For the full-$fp$-shell calculation, we use the FPD6 interaction~\cite{rmjb91}.
The interaction obtained by setting the $T=0$ matrix elements to zero but 
keeping the $T=1$ ones unchanged is called T0FPD6.

We list the results in Table~\ref{tab:full} for the full-$fp$-shell 
calculation. Note that with T0FPD6 the levels in question are no longer 
degenerate, but the splittings are, for the most part, much larger when the 
full FPD6 interaction is turned on.

\begin{table}[ht]
\caption{Full-$fp$-shell excitation energies (MeV) for states that were 
degenerate in the single $j$ shell with $T0E(J)$. All experimental energies are
taken from Ref.~\cite{nndc}.} 
\label{tab:full}
\begin{tabular*}{.8\textwidth}{@{\extracolsep{\fill}}ccddd}
\toprule
 & $J$ & \multicolumn{1}{c}{FPD6} & \multicolumn{1}{c}{T0FPD6} & 
\multicolumn{1}{c}{Experiment} \\ \colrule
$^{43}$Sc ($^{43}$Ti) & $(1/2)^-_1$ & 1.809 & 2.915 & \\
 & $(13/2)^-_1$ & 3.675 & 3.041  \\
 & $(13/2)^-_2$ & 4.779 & 3.648 \\
 & $(17/2)^-_1$ & 4.380 & 3.671 & 4.360 \\
 & $(19/2)^-_1$ & 3.360 & 3.581 & 3.123 \\ 
$^{44}$Ti & $3^+_2$ & 8.176 & 5.531 \\
 & $7^+_2$ & 9.207 & 6.635 \\
 & $9^+_1$ & 8.828 & 6.454 \\
 & $10^+_1$ & 7.614 & 6.360 & 7.671 \\
 & $10^+_2$ & 9.929 & 7.194 & 8.984\footnotemark[1] \\
 & $12^+_1$ & 8.312 & 7.061 & 8.039 \\ 
$^{45}$Ti & $(25/2)^-_1$ & 8.652 & 6.955 & \\ 
 & $(27/2)^-_1$ & 7.697 & 6.850 & 7.143 \\ 
$^{46}$V & $12^+_1$ & 7.841 & 8.276 & 8.268 \\
 & $12^+_2$ & 8.729 & 8.669 \\
 & $13^+_1$ & 7.341 & 8.106 & 7.105 \\
 & $13^+_2$ & 10.389 & 9.735 \\
 & $15^+_1$ & 8.995 & 9.559 & 8.488 \\ 
$^{47}$V & $(29/2)^-_1$ & 11.848 & 9.412 & 10.7685\footnotemark[1] \\
 & $(31/2)^-_1$ & 11.068 & 9.340 & 10.003
\\
\botrule
\end{tabular*}
\footnotetext[1]{Taken from the Experimental Unevaluated Nuclear Data List.}
\end{table}

\begin{table}[ht]
\caption{Splitting in energies (MeV) for states that were 
degenerate in the single $j$ shell with $T0E(J)$. All experimental energies are
taken from Ref.~\cite{nndc}.} \label{tab:split}
\begin{tabular*}{.9\textwidth}{@{\extracolsep{\fill}}ccdddd}
\toprule
 & $\Delta E$ & \multicolumn{1}{c}{Single $j$} & \multicolumn{1}{c}{FPD6} & 
\multicolumn{1}{c}{T0FPD6} & \multicolumn{1}{c}{Exp.} \\ \colrule
$^{43}$Sc ($^{43}$Ti) &  $(1/2)_1^- -(13/2)^-_1$ & 0.816 & -1.866 & -0.126 \\
 & $(13/2)_2^- -(17/2)_1^-$ & 0.653 & 0.399 & -0.023 \\
 & $(17/2)_1^- -(19/2)_1^-$ & 0.653 & 1.020 & 0.090 & 1.237 \\
$^{44}$Ti & $3_2^+ -7_2^+$ & 0.320 & -1.031 & -1.104 \\
 & $7_2^+ -9_1^+$ & 0.391 & 0.379 & 0.181 \\
 & $9_1^+ -10_1^+$ & 0.600 & 1.214 & 0.094 \\
 & $10_2^+ -12_1^+$ & 1.203 & 1.617 & 0.133 & 0.945 \\
$^{45}$Ti & $(25/2)_1^- -(27/2)_1^-$ & 0.580 & 0.955 & 0.105 & \\ 
$^{46}$V & $12_1^+ -13_1^+$ & 0.863 & 0.500 & 0.170 & 1.163 \\
 & $13_2^+ -15_1^+$ & 0.809 & 1.394 & 0.176 \\
$^{47}$V & $(29/2)_1^- -(31/2)_1^-$ & 0.229 & 0.780 & 0.072 & 0.765 \\
\botrule
\end{tabular*}
\end{table}

As a comparison, we give in Table~\ref{tab:split} the results for the shifts 
$\Delta E$ in both the single-$j$-shell case and the full $fp$ calculation.
With two exceptions, the results in Table~\ref{tab:split} show a continuity 
between single $j$ and the full $fp$ calculation. The exceptions, $(1/2^-_1 -
13/2^-_1)$ in $^{43}$Sc and $(3^+_2 -7^+_2)$ in $^{44}$Ti, will soon be 
discussed.

Note that, aside from the above exceptions, the shifts $\Delta E$ are much 
smaller for T0FPD6 than for FPD6. For example, the $(9^+_1 -10^+_1)$ splitting 
is 1.214~MeV for FPD6, but it is only 0.094~MeV for T0FPD6. This supports the 
fact that most of the splitting comes from the $T=0$ part of the two-body 
interaction.

The two exceptions mentioned above seem to occur for low angular momentum 
states: $1/2^-$ in $^{43}$Sc and $3^+$ in $^{44}$Ti. For these low-lying 
states, there tends to be much more configuration mixing; hence, some states 
having very little to do with the $f_{7/2}$ configuration must have slipped 
down in energy.


We have previously studied the effects of removing and then reinserting the
$T=0$ two-body matrix elements in nuclear structure calculations~\cite{rez05,
rz02}. In Ref.~\cite{rez05} we compared the yrast spectra of even--even nuclei
in the $fp$ shell using FPD6 and T0FPD6. While the reintroduction of the $T=0$ 
matrix elements causes the spectra to become more rotational, it is clear that 
the $T=1$ part of the interaction dominates the spectrum. Thus, it is not easy 
to get a quantitative handle on the effects of the $T=0$ two-body matrix 
elements by looking at the yrast spectrum alone.

Therefore, we are looking for various benchmarks that, in combination, will 
help to obtain the correct $T=0$ effective interaction. In Ref.~\cite{rez05} it
was noted that the $B(E2)$'s were enhanced by the reintroduction of $T=0$ 
matrix elements, but this point is somewhat obscured by the uncertainty in what
effective charges should be used for neutrons and protons. 

In Ref.~\cite{rz02} it was noted that, in some channels, the Gamow-Teller (GT)
matrix elements were very sensitive to the $T=0$ interaction and, perhaps most 
important, the isovector orbital transition strength (scissors mode) was 
greatly enhanced with FPD6 relative to T0FPD6.

In this work, we have focussed on measurements and calculations where the $T=0$
two-body matrix elements are very important for obtaining energy splittings of 
states that would be degenerate in single-$j$-shell calculations using T0FPD6. 
Although there are some cases where the single-$j$ calculations are closer to 
experiment than the full calculation with FPD6, it is clear that the study of 
the correct $T=0$ interaction will have to be carried out using a full $fp$ 
space.

From the results in Table~\ref{tab:split}, it appears that the high-spin 
splittings are the simplest to put to the test. Note, however, that some 
experimental data are missing, e.g., the $(25/2^-_1 - 27/2^-_1)$ splitting in 
$^{45}$Ti and $(13^+_2 - 15^+_1)$ in $^{46}$V.

We would like to acknowledge support from a U.S. Dept. of Energy Grant No. 
DE-FG0105ER05-02 and from the Secretar\'{\i}a de Estado de Educaci\'on y 
Universidades (Spain) and the European Social Fund. We thank Ben Bayman for
technical help and stimulating discussions.

\end{document}